\documentstyle[aps,pre,twocolumn,epsfig,tabularx]{revtex}
\newcommand{\bel}[1]{\begin{equation}\label{#1}}
\newcommand{\bra}[1]{\mbox{$\langle \, {#1}\, |$}}
\newcommand{\ket}[1]{\mbox{$| \, {#1}\, \rangle$}}
\newcommand{\exval}[1]{\mbox{$\langle \, {#1}\, \rangle$}}
\def\bbbc{{\mathchoice {\setbox0=\hbox{$\displaystyle\rm C$}\hbox{\hbox
to0pt{\kern0.4\wd0\vrule height0.9\ht0\hss}\box0}}
{\setbox0=\hbox{$\textstyle\rm C$}\hbox{\hbox
to0pt{\kern0.4\wd0\vrule height0.9\ht0\hss}\box0}}
{\setbox0=\hbox{$\scriptstyle\rm C$}\hbox{\hbox
to0pt{\kern0.4\wd0\vrule height0.9\ht0\hss}\box0}}
{\setbox0=\hbox{$\scriptscriptstyle\rm C$}\hbox{\hbox
to0pt{\kern0.4\wd0\vrule height0.9\ht0\hss}\box0}}}}
\def\be{\begin{equation}}
\def\ee{\end{equation}}
\def\bea{\begin{eqnarray}}
\def\eea{\end{eqnarray}}
\def\ba{\begin{array}}
\def\ea{\end{array}}
\draft
\begin{document}
\tighten
\onecolumn
\twocolumn[\hsize\textwidth\columnwidth\hsize\csname @twocolumnfalse\endcsname

\title{Asymmetric exclusion process with next-nearest-neighbor interaction: 
some comments on traffic flow and a nonequilibrium reentrance transition
}
\date{\today}
\author{T. Antal$^{1,2}$ and G. M. Sch\"utz$^{2}$}
\address{$^1$ D\'epartement de Physique Th\'eorique, Universit\'e de Gen\`eve, 
1211 Gen\`eve 4, 24 quai Ernest Ansermet, Switzerland\\
$^2$ Institut f\"ur Festk\"orperforschung, Forschungszentrum J\"ulich,
52425 J\"ulich, Germany}
\maketitle
\begin{abstract}
We study the steady-state behavior of a driven non-equilibrium lattice gas
of hard-core particles with next-nearest-neighbor interaction. We calculate the
exact stationary distribution of the periodic system and for a particular line
in the phase diagram of the system with open boundaries where particles
can enter and leave the system. For repulsive interactions the dynamics can 
be interpreted as a two-speed model for traffic flow. 
The exact stationary distribution of the 
periodic continuous-time system turns out to coincide with that of the 
asymmetric exclusion process (ASEP) with discrete-time parallel update. 
However, unlike in the (single-speed) ASEP, the exact flow diagram for the 
two-speed model resembles in some important features the flow diagram of real 
traffic. The stationary phase diagram of the open system obtained from Monte 
Carlo simulations can be understood in terms of a shock moving through the 
system and an overfeeding effect at the boundaries, thus
confirming theoretical predictions of a recently developed general theory of
boundary-induced phase transitions. In the case of attractive interaction
we observe an unexpected reentrance transition due to boundary effects.
\end{abstract}
\pacs{PACS: 05.70.Ln, 64.60.Cn, 02.50.Ga}
]
\section{Introduction}
\label{Intro}

One of the basic properties of many driven, interacting many-body systems
is the occurrence of shocks. A shock in a system of classical flowing particles
marks the sudden transition from a region of low density to a region of high
density. A well-known example for a shock is the beginning of a traffic jam on
a motorway where incoming cars (almost freely flowing particles in the low
density regime) have to slow down very quickly over a short distance and then
form part of the (high-density) congested region. A remarkable feature of such
shocks is their stability over long periods of time, i.e., they remain
localized over distances comparable to the size of particles. In some sense
one may regard shocks as soliton-like collective excitations of the particle
system.

Lattice gas models have proven to be excellent systems for the theoretical
investigation of shocks and of the consequences of shocks for the collective
behavior of the particle system. An interesting situation is the coupling of
such a driven lattice gas system to external particle reservoirs
\cite{MacD68,Ligg75,Krug91a}. It is intuitively clear that unlike
in equilibrium systems, here the boundaries will play a decisive part in
determining the bulk behavior of the system: Since the system is open
at the boundaries, particles will flow in and out and the current will
carry boundary effects into the bulk. Among the
fundamental questions to ask in such a set-up is the stationary
(i.e. long-time) behavior of the system as a function of the boundary
densities. Numerical observations and mean-field based arguments show that
varying boundary densities leads to boundary-induced phase transitions
\cite{Krug91a}. To understand why this happens it is clearly necessary to get
insight into the collective behavior of the lattice gas and to investigate
the role of the shocks.

The best-studied example is the one-dimensional asymmetric simple exclusion
process (ASEP) \cite{Ligg85,Schu00,Priv97}, a hard-core lattice gas where each 
lattice site can be occupied by at most one particle and particles hop 
stochastically with constant bias to vacant nearest-neighbor sites. For this 
model not only the structure and motion of shocks is largely understood, but 
also the stationary phase diagram of the open system \cite{Ligg75} and the exact 
particle-density profiles \cite{Schu93b,Derr93a} are explicitly known as 
function of the external reservoir densities. Based on the exact solution of 
the ASEP and of a related lattice gas model \cite{Schu93c} the role of shocks 
for the stationary phase of more generic open driven diffusive systems was 
elucidated \cite{Kolo98,Popk99a}. This has led to a theory of boundary-induced 
phase transitions for one-component driven particle systems in one dimension
\cite{Schu00}, reviewed below. Of particular interest are systems where the 
stationary current $j(\rho)$ as a function of particle density $\rho$ has a 
single maximum, an example being traffic flow on single-lane or multi-lane 
highways \cite{Chow00,Helb97}. The theory predicts a phase diagram with
a first-order (non-equilibrium) phase transition between a low-density phase
(LD) and a high density phase (HD) and a continuous (second-order) order
transition from both phases to a maximal-current phase (MC), see Fig.~2
below. In the context of traffic flow the properties of the second-order
transition suggest more efficient control mechanisms for avoiding jams 
\cite{Popk99b}.

The strength of the theory lies in the small number and generality of its
assumptions. Hence it is tempting to test its validity in real traffic
flow. The openness of the system models in- and outflow of cars on a road 
between two junctions. Indeed, the main features of the predicted first-order 
transition have been observed using data collected on a German motorway 
\cite{Popk99b,Neub99}. The second-order transition has been confirmed by 
Monte-Carlo simulations of a suitably modified Nagel-Schreckenberg model for 
traffic flow, originally introduced only for periodic boundary conditions 
\cite{Nage92,Scha93,Schr95}. Since to 
our knowledge the data necessary to test the existence of the second-order
transition in real traffic are not available at present, independent
investigations of other traffic flow models are required to establish
the presence of such a transition. 
We consider here a continuous-time exclusion model with additional short-range 
interaction on top of the pure hard-core repulsion of the usual ASEP. The model
is designed to be exactly solvable like the ASEP (to some degree) on the 
one hand and to be somewhat more realistic in describing real traffic than
the ASEP on the 
other hand. However, given the wide range of experimental
applications of hard-core lattice gases which comprise phenomena
as diverse as the kinetics of protein synthesis \cite{MacD68}, diffusion in
thin channels \cite{Kukl96}, or polymer reptation \cite{Bark96}, broader
relevance of our model is to be expected. Indeed, in order to obtain a more
complete picture we also analyse a similar model with attractive 
next-nearest-neighbour interaction. This model is not related to traffic flow, 
but describes driven hard-core particles with attractive short-range interaction.
To our knowledge, this is the first investigation of the steady-state selection 
of a driven diffusive system with this type of interaction.

The paper is organized as follows. In Sec. \ref{curr} we comment on some of the
requirements of traffic flow modeling and we describe our model. This model
is a special case of the driven diffusive systems studied some
while ago by Katz, Lebowitz and Spohn \cite{Katz84}. We explain in which
respect this model is more realistic for traffic flow than the standard ASEP
which is a limiting case of our lattice gas. This phenomenological explanation
is then confirmed by the calculation of the exact flow diagram,
i.e. the stationary bulk current as a function of the density. In this section
the emphasis is on bulk properties and hence we investigate the system
with periodic boundary conditions in the  thermodynamic limit.
In Sec. \ref{open} we review some details of the
theory of boundary-induced phase transitions of Ref. \cite{Kolo98} and
we introduce the boundary dynamics for modeling the coupling of a finite system
to boundary reservoirs of constant density. We give an exact solution
for the stationary distribution along the line in the phase diagram
corresponding to equal boundary densities (with proof postponed to the 
appendices) and discuss a mean-field analysis of the full phase diagram.
In Sec. \ref{rep} we present a mean-field analysis of the full phase diagram
and discuss Monte Carlo data for the phase transition lines while in Sec.
\ref{att} we perform a similar analysis for the model with attractive
interaction. Finally we summarize our findings and present some
concluding remarks (Sec. \ref{conc}). Technical details of the derivation
of exact results are presented in the appendices.

\section{ASEP with next-nearest-neighbor interaction - some comments on 
modelling traffic flow}
\label{curr}

The ASEP is a stochastic lattice gas of hard-core particles with
biased particle hopping. Particles hop with exponential
waiting-time distribution with parameter 1 to their nearest-neighbor
site to the right, provided this site is empty. If the site is occupied,
the hopping attempt is rejected. Symbolically one may represent these
stochastic dynamics as follows
\bea
A \emptyset & \to & \emptyset A \hspace{2mm} \mbox{ with rate } 1.
\eea
Here $A$ represent a particle and $\emptyset$ represents the vacant
neighboring site. Even though this stochastic process is too simple to be a
realistic model for traffic flow, some {\em qualitative} features of real
traffic \cite{Hall86,Kern96} can already be seen: Shocks exist and the 
stationary current $j(\rho)=\rho(1-\rho)$ as a function of particle density 
$\rho$ has a single maximum. An apparently unrealistic feature of the particle
distribution is the absence of correlations in the steady state which are
seen in more sophisticated traffic flow models
\cite{Scha93,Schr95}. An unrealistic feature of the current-density relation
(known in traffic engineering as flow diagram) is the reflection symmetry
w.r.t. the maximal-current density $\rho^\ast = 1/2$ and its rounded shape
close to the maximum.

The basic mechanisms which determine traffic flow appear to be firstly the 
competition between the desire to reach an optimal speed while keeping a 
(velocity-dependent) safety distance. When the traffic density becomes 
sufficiently large this distance cannot be kept anymore and the speed has to 
be reduced. As a consequence the current drops at some density $\rho^\ast$. 
Secondly, there is a certain amount of randomness due to variations in 
individual drivers behavior. This ``noise'' necessitates
a statistical description of traffic flow phenomena. 

Various one-dimensional lattice gas models which incorporate these mechanisms in 
different manners have been proposed \cite{Chow00}. The following is part of the 
picture that emerges:\\
(i) The existence of a shock is generic and appears to be the 
consequence of the non-linear current-density relation \cite{Schu00}.\\ 
(ii) The symmetric shape 
of the current-density relation results from particle-hole symmetry and is a 
feature of models in which cars move with constant probability or rate,
independently of the environment beyond the nearest neighbor site to which
they move. This is an unrealistic assumption since clearly car drivers slow down
when they see a slowly moving car already some distance ahead. They do
not just perform an emergency stop when the car is immediately in front
of them. Numerical and analytical results for models \cite{Nage92,Schr95,Klau99} 
which allow for a reduction of speed that depends on the occupation of sites 
further ahead show an asymmetric current-density relation resembling the shape
of the current-density relation of real traffic. In these models speed
is implemented by jumps over a variable number of lattice sites.\\
(iii) The unrealistically round shape of the current-density relation at 
$\rho^\ast$ is specific for the ASEP. Deterministic exclusion processes with 
parallel update \cite{Krug88,Schu93a,Yuka94,Raje97,Tils98,deGi99}
also show a symmetric current-density relation with one maximum,
but the derivative of the current is discontinuous at the maximal-current
density $\rho^\ast$, in this respect resembling the shape of the current in
real traffic \cite{Hall86} and of more realistic traffic flow models 
\cite{Nage92}. Increasing the hopping probability in a discrete-time process 
towards deterministic hopping, leads to an increasingly sharp jump in the 
current derivative at $\rho^\ast$ \cite{Scha93}. Hence the rather broad 
exponential waiting-time distribution of the standard ASEP seems to be 
responsible for the round shape of the current-density relation at $\rho^\ast$.
In terms of the motion of a single particle these dynamics correspond to
an overestimated single-particle diffusion coefficient.\\ 
(iv) For parallel update, but not for sublattice parallel update, increasing 
the hopping probability strengthens antiferromagnetic particle correlations 
\cite{Schr95,Raje97}, i.e., cars are less likely to be found close to each
other than some distance apart. 

In order to further investigate this picture and to disentangle the various 
effects associated with a small diffusion coefficient and with speed-reduction 
respectively it would be interesting to study both ingredients separately for 
models in which the stationary distribution can be calculated exactly. A small 
single-particle diffusion coefficient can be implemented in discrete-time 
models choosing the hopping probability close to one (low noise). In the 
single-speed ASEP this has been shown to lead to the (almost) discontinuous 
behavior of the current derivative discussed above. The numerical results on 
the effect of a small diffusion coefficient on correlations are inconclusive. 
In our toy model we keep the exponential waiting-time distribution (large 
diffusion coefficient), but introduce a next-nearest-neighbor interaction which 
in the repulsive case models slowing down of a car if the next-nearest-neighbor 
site is occupied as well. A particle hops to the right with rate $r$ if the
next-nearest-neighbor site is empty and with rate $q$ if it is occupied:
\bea
A \emptyset\emptyset & \to & \emptyset A \emptyset\hspace{1cm}
\mbox{ with rate } r \\
A \emptyset A & \to & \emptyset A A\hspace{1cm}
\mbox{ with rate } q.
\eea
The condition $q<r$ models slowing-down, in the limiting case $r=q$
one recovers the usual ASEP. For $q>r$ this model has not an interpretation as 
traffic model, but may be regarded as describing hard-core particles with 
attractive short-range interaction which are driven by an external field.

This model is a special case in the class of driven diffusive systems
investigated in Ref. \cite{Katz84}. On a ring with $N$ sites with
periodic boundary conditions the stationary distribution turns out to
be given by the equilibrium distribution of the one-dimensional
Ising model. Each state of the system is defined by the set of occupation
numbers $\underline{n} = \{n_1,\dots,n_N\}$ with $n_i = 0,1$. The stationary
probability of finding a state $\underline{n}$ is given by
\bel{2-1}
P^\ast( \underline{n}) = \frac{1}{Z_N}
\left(\frac{q}{r}\right)^{\sum_{i=1}^N (n_i n_{i+1} +  h n_i)}.
\ee
Here $Z_N$ is the partition function and the ``chemical potential'' $h$
parametrizes the conserved bulk density $\rho$. This grand-canonical
distribution is a non-equilibrium stationary state, i.e., it is invariant under 
the stochastic time evolution, but it does not satisfy detailed balance with 
respect to the dynamics.
It is interesting to notice that correlations are non-vanishing and, in the
repulsive case $q<r$ which corresponds to speed reduction, become
antiferromagnetic. In fact, the stationary state is identical
to that of the discrete-time ASEP with parallel update \cite{Schr95,Yagu86}
for hopping probability $p=1-q/r$.

\begin{figure}[htb]
\centerline{
\epsfxsize=0.4\textwidth
\epsfbox{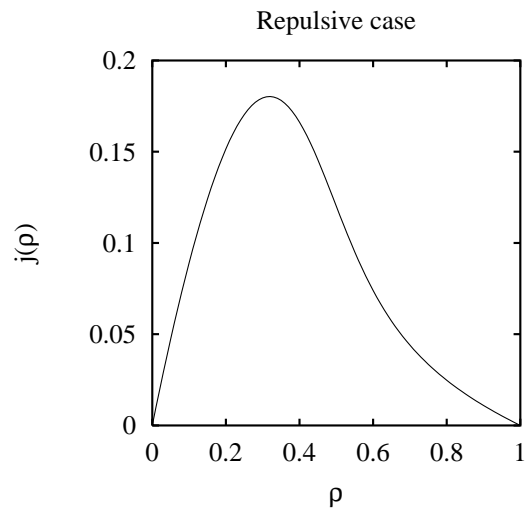}}
\caption{Stationary current $j$ as a function of the density $\rho$
for $r=1,\, q=0.1$ (repulsive interaction).}
\label{AS98F-1}
\end{figure}

According to the dynamics described above the local density satisfies the
continuity equation
\bel{2-1a}
\frac{d}{dt} \exval{n_i} = \exval{j_{i-1}} - \exval{j_{i}}
\ee
with the current
\bel{2-2}
\exval{j_{i}} = \exval{n_i(1-n_{i+1})[q n_{i+2}+r(1-n_{i+2})]}
\ee
between sites $(i,i+1)$.
The stationary particle current $j = \exval{j_{i}}$ is constant and
is readily calculated
using standard transfer matrix techniques for the one-dimensional Ising model
\cite{Baxt82} (see Appendix A). In the thermodynamic limit $N \to \infty$
one finds the exact current density relation
\bel{2-3}
j = r \rho \left[1 +
\frac{\sqrt{1-4\rho(1-\rho)(1-q/r)}-1}{2(1-\rho)(1-q/r)} \right] 
\ee
shown in (Fig.~\ref{AS98F-1}) in the repulsive case.

\section{Open boundaries}
\label{open}

The analysis of the previous section yields the bulk properties of the
lattice gas at a given density. Now we address the question of the
steady-state selection of the system with open boundaries, i.e., we
investigate the bulk density of an open system coupled at both ends to 
reservoirs of different fixed boundary densities. Translated
into traffic language, open boundaries correspond to traffic junctions
at the ends of a road where particles (cars) enter or leave
the road respectively with certain fixed attempt rates,
a situation envisaged with different aims already in \cite{Naga95}
for the ASEP with parallel update. As a result of the
coupling to
reservoirs a non-trivial stationary density profile in the vicinity of the
boundaries will emerge and the (spatially constant) bulk-density will be a 
function of the two reservoir densities.

\begin{figure}[htbp]
\setlength{\unitlength}{7mm}
\epsfig{width=3.8\unitlength, angle=-90,
      file=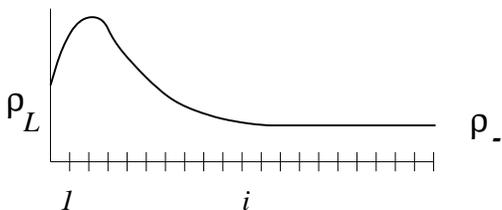}\\
\caption[]{Coupling of a semi-infinite driven particle system to a reservoir
of constant particle density $\rho_L$. The (interpolated) density profile 
$\rho_i$ ($i=1,2,\dots$) approaches some constant value $\rho_-$ after some 
finite distance from the boundary.}
\label{bound}
\end{figure}

There are two distinct mechanisms at work. Firstly, because of the particle
interaction, coupling of a semi-infinite system to a reservoir will
generically lead to some discontinuous behavior of the stationary
distribution close to the boundary. The boundary represents an inhomogeneity
of the system since the interaction of the particles with the fixed boundary
leads to different dynamics than that which results from the interaction of
particles among themselves. This is a non-universal phenomenon which depends
on the precise nature of the coupling mechanism and on the nature of the
particle interaction. For short-range
hopping and systems with short-range stationary bulk correlations one expects
the following picture: Coupling of a semi-infinite system at site 1 to a
reservoir of constant density $\rho_L$ will give rise to a non-universal
boundary density profile starting at $\rho_L$ and approaching
(on the scale of lattice units) some
bulk density $\rho_{-}$ which is a non-universal function of $\rho_L$
(Fig.~\ref{bound}).
A similar picture holds for coupling of a semi-infinite system at the
right boundary where particles flow out of the system into the reservoir.
Here the bulk density $\rho_{+}$ may change close to the boundary to
the reservoir density $\rho_R$.
Superimposed on this non-universal boundary structure is a universal behavior
which depends only the effective boundary densities $\rho_{-},\rho_{+}$
close to,
but not at the boundaries. The theory of boundary induced phase transition
\cite{Kolo98} describes the stationary phase diagram in terms of these
effective boundary densities.

\subsection{Theory of boundary-induced phase transitions}

We review only the principal ideas of
this theory \cite{Schu00} which is based on an interplay of the
collective velocity
\bel{3-1}
v_c = \frac{\partial j}{\partial \rho}
\ee
of the lattice gas and the shock velocity
\bel{3-2}
v_s = \frac{j_{+} - j_-}{\rho_{+} - \rho_{-}}
\ee
of a shock with limiting densities $\rho_{+}$ and $\rho_{-}$ 
and with limiting currents $j_+$ and $j_-$ to the right and to the left respectively.

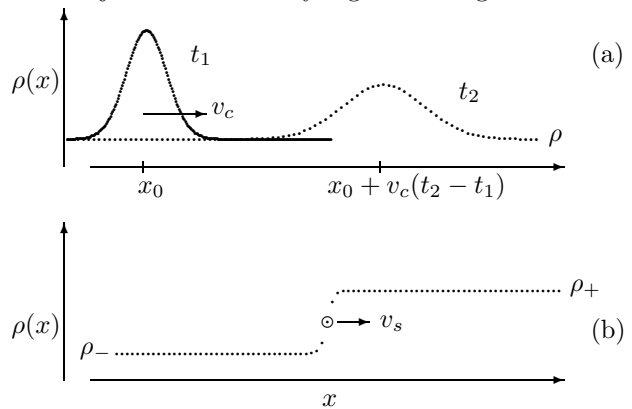
\begin{figure}
\setlength{\unitlength}{0.7mm}
\begin{center}
\begin{picture}(100,30)(0,0)

\put(100,25){(a)}

\put(0,5){\vector(0,1){30}}
\put(-10,20){$\rho(x)$}
\put(92,10){$\rho$}

\put(5,5){\vector(1,0){90}}
\put(14,0){$x_0$}
\put(15,4){\line(0,1){2}}
\put(50,0){$x_0 + v_c(t_2 - t_1)$}
\put(60,4){\line(0,1){2}}

\put(24,25){$t_1$}
\put(75,18){$t_2$}
\put(15,15){\vector(1,0){12}}
\put(28,15){$v_c$}

\put(0,10.){.}
\put(0.2,10.){.}
\put(0.4,10.){.}
\put(0.6,10.){.}
\put(0.8,10.){.}
\put(1.,10.){.}
\put(1.2,10.){.}
\put(1.4,10.){.}
\put(1.6,10.1){.}
\put(1.8,10.1){.}
\put(2.,10.1){.}
\put(2.2,10.1){.}
\put(2.4,10.1){.}
\put(2.6,10.1){.}
\put(2.8,10.1){.}
\put(3.,10.2){.}
\put(3.2,10.2){.}
\put(3.4,10.2){.}
\put(3.6,10.3){.}
\put(3.8,10.3){.}
\put(4.,10.4){.}
\put(4.2,10.4){.}
\put(4.4,10.5){.}
\put(4.6,10.6){.}
\put(4.8,10.6){.}
\put(5.,10.7){.}
\put(5.2,10.8){.}
\put(5.4,11.){.}
\put(5.6,11.1){.}
\put(5.8,11.2){.}
\put(6.,11.4){.}
\put(6.2,11.6){.}
\put(6.4,11.8){.}
\put(6.6,12.){.}
\put(6.8,12.2){.}
\put(7.,12.4){.}
\put(7.2,12.7){.}
\put(7.4,13.){.}
\put(7.6,13.3){.}
\put(7.8,13.7){.}
\put(8.,14.){.}
\put(8.2,14.4){.}
\put(8.4,14.8){.}
\put(8.6,15.3){.}
\put(8.8,15.7){.}
\put(9.,16.2){.}
\put(9.2,16.7){.}
\put(9.4,17.2){.}
\put(9.6,17.8){.}
\put(9.8,18.4){.}
\put(10.,19.){.}
\put(10.2,19.6){.}
\put(10.4,20.2){.}
\put(10.6,20.8){.}
\put(10.8,21.4){.}
\put(11.,22.1){.}
\put(11.2,22.7){.}
\put(11.4,23.4){.}
\put(11.6,24.){.}
\put(11.8,24.6){.}
\put(12.,25.3){.}
\put(12.2,25.9){.}
\put(12.4,26.4){.}
\put(12.6,27.){.}
\put(12.8,27.5){.}
\put(13.,28.){.}
\put(13.2,28.5){.}
\put(13.4,28.9){.}
\put(13.6,29.3){.}
\put(13.8,29.6){.}
\put(14.,29.9){.}
\put(14.2,30.2){.}
\put(14.4,30.4){.}
\put(14.6,30.5){.}
\put(14.8,30.6){.}
\put(15.,30.6){.}
\put(15.2,30.6){.}
\put(15.4,30.5){.}
\put(15.6,30.4){.}
\put(15.8,30.2){.}
\put(16.,29.9){.}
\put(16.2,29.6){.}
\put(16.4,29.3){.}
\put(16.6,28.9){.}
\put(16.8,28.5){.}
\put(17.,28.){.}
\put(17.2,27.5){.}
\put(17.4,27.){.}
\put(17.6,26.4){.}
\put(17.8,25.9){.}
\put(18.,25.3){.}
\put(18.2,24.6){.}
\put(18.4,24.){.}
\put(18.6,23.4){.}
\put(18.8,22.7){.}
\put(19.,22.1){.}
\put(19.2,21.4){.}
\put(19.4,20.8){.}
\put(19.6,20.2){.}
\put(19.8,19.6){.}
\put(20.,19.){.}
\put(20.2,18.4){.}
\put(20.4,17.8){.}
\put(20.6,17.2){.}
\put(20.8,16.7){.}
\put(21.,16.2){.}
\put(21.2,15.7){.}
\put(21.4,15.3){.}
\put(21.6,14.8){.}
\put(21.8,14.4){.}
\put(22.,14.){.}
\put(22.2,13.7){.}
\put(22.4,13.3){.}
\put(22.6,13.){.}
\put(22.8,12.7){.}
\put(23.,12.4){.}
\put(23.2,12.2){.}
\put(23.4,12.){.}
\put(23.6,11.8){.}
\put(23.8,11.6){.}
\put(24.,11.4){.}
\put(24.2,11.2){.}
\put(24.4,11.1){.}
\put(24.6,11.){.}
\put(24.8,10.8){.}
\put(25.,10.7){.}
\put(25.2,10.6){.}
\put(25.4,10.6){.}
\put(25.6,10.5){.}
\put(25.8,10.4){.}
\put(26.,10.4){.}
\put(26.2,10.3){.}
\put(26.4,10.3){.}
\put(26.6,10.2){.}
\put(26.8,10.2){.}
\put(27.,10.2){.}
\put(27.2,10.1){.}
\put(27.4,10.1){.}
\put(27.6,10.1){.}
\put(27.8,10.1){.}
\put(28.,10.1){.}
\put(28.2,10.1){.}
\put(28.4,10.1){.}
\put(28.6,10.){.}
\put(28.8,10.){.}
\put(29.,10.){.}
\put(29.2,10.){.}
\put(29.4,10.){.}
\put(29.6,10.){.}
\put(29.8,10.){.}
\put(30.,10.){.}
\put(30.2,10.){.}
\put(30.4,10.){.}
\put(30.6,10.){.}
\put(30.8,10.){.}
\put(31.,10.){.}
\put(31.2,10.){.}
\put(31.4,10.){.}
\put(31.6,10.){.}
\put(31.8,10.){.}
\put(32.,10.){.}
\put(32.2,10.){.}
\put(32.4,10.){.}
\put(32.6,10.){.}
\put(32.8,10.){.}
\put(33.,10.){.}
\put(33.2,10.){.}
\put(33.4,10.){.}
\put(33.6,10.){.}
\put(33.8,10.){.}
\put(34.,10.){.}
\put(34.2,10.){.}
\put(34.4,10.){.}
\put(34.6,10.){.}
\put(34.8,10.){.}
\put(35.,10.){.}
\put(35.2,10.){.}
\put(35.4,10.){.}
\put(35.6,10.){.}
\put(35.8,10.){.}
\put(36.,10.){.}
\put(36.2,10.){.}
\put(36.4,10.){.}
\put(36.6,10.){.}
\put(36.8,10.){.}
\put(37.,10.){.}
\put(37.2,10.){.}
\put(37.4,10.){.}
\put(37.6,10.){.}
\put(37.8,10.){.}
\put(38.,10.){.}
\put(38.2,10.){.}
\put(38.4,10.){.}
\put(38.6,10.){.}
\put(38.8,10.){.}
\put(39.,10.){.}
\put(39.2,10.){.}
\put(39.4,10.){.}
\put(39.6,10.){.}
\put(39.8,10.){.}
\put(40.,10.){.}
\put(40.2,10.){.}
\put(40.4,10.){.}
\put(40.6,10.){.}
\put(40.8,10.){.}
\put(41.,10.){.}
\put(41.2,10.){.}
\put(41.4,10.){.}
\put(41.6,10.){.}
\put(41.8,10.){.}
\put(42.,10.){.}
\put(42.2,10.){.}
\put(42.4,10.){.}
\put(42.6,10.){.}
\put(42.8,10.){.}
\put(43.,10.){.}
\put(43.2,10.){.}
\put(43.4,10.){.}
\put(43.6,10.){.}
\put(43.8,10.){.}
\put(44.,10.){.}
\put(44.2,10.){.}
\put(44.4,10.){.}
\put(44.6,10.){.}
\put(44.8,10.){.}
\put(45.,10.){.}
\put(45.2,10.){.}
\put(45.4,10.){.}
\put(45.6,10.){.}
\put(45.8,10.){.}
\put(46.,10.){.}
\put(46.2,10.){.}
\put(46.4,10.){.}
\put(46.6,10.){.}
\put(46.8,10.){.}
\put(47.,10.){.}
\put(47.2,10.){.}
\put(47.4,10.){.}
\put(47.6,10.){.}
\put(47.8,10.){.}
\put(48.,10.){.}
\put(48.2,10.){.}
\put(48.4,10.){.}
\put(48.6,10.){.}
\put(48.8,10.){.}
\put(49.,10.){.}
\put(49.2,10.){.}
\put(49.4,10.){.}
\put(49.6,10.){.}
\put(49.8,10.){.}
\put(50.,10.){.}
\put(0,10.){.}
\put(1.,10.){.}
\put(2.,10.){.}
\put(3.,10.){.}
\put(4.,10.){.}
\put(5.,10.){.}
\put(6.,10.){.}
\put(7.,10.){.}
\put(8.,10.){.}
\put(9.,10.){.}
\put(10.,10.){.}
\put(11.,10.){.}
\put(12.,10.){.}
\put(13.,10.){.}
\put(14.,10.){.}
\put(15.,10.){.}
\put(16.,10.){.}
\put(17.,10.){.}
\put(18.,10.){.}
\put(19.,10.){.}
\put(20.,10.){.}
\put(21.,10.){.}
\put(22.,10.){.}
\put(23.,10.){.}
\put(24.,10.){.}
\put(25.,10.){.}
\put(26.,10.){.}
\put(27.,10.){.}
\put(28.,10.){.}
\put(29.,10.){.}
\put(30.,10.){.}
\put(31.,10.){.}
\put(32.,10.){.}
\put(33.,10.){.}
\put(34.,10.){.}
\put(35.,10.1){.}
\put(36.,10.1){.}
\put(37.,10.1){.}
\put(38.,10.2){.}
\put(39.,10.3){.}
\put(40.,10.4){.}
\put(41.,10.5){.}
\put(42.,10.7){.}
\put(43.,10.9){.}
\put(44.,11.2){.}
\put(45.,11.6){.}
\put(46.,12.){.}
\put(47.,12.5){.}
\put(48.,13.1){.}
\put(49.,13.8){.}
\put(50.,14.5){.}
\put(51.,15.2){.}
\put(52.,16.){.}
\put(53.,16.8){.}
\put(54.,17.6){.}
\put(55.,18.4){.}
\put(56.,19.){.}
\put(57.,19.6){.}
\put(58.,20.){.}
\put(59.,20.2){.}
\put(60.,20.3){.}
\put(61.,20.2){.}
\put(62.,20.){.}
\put(63.,19.6){.}
\put(64.,19.){.}
\put(65.,18.4){.}
\put(66.,17.6){.}
\put(67.,16.8){.}
\put(68.,16.){.}
\put(69.,15.2){.}
\put(70.,14.5){.}
\put(71.,13.8){.}
\put(72.,13.1){.}
\put(73.,12.5){.}
\put(74.,12.){.}
\put(75.,11.6){.}
\put(76.,11.2){.}
\put(77.,10.9){.}
\put(78.,10.7){.}
\put(79.,10.5){.}
\put(80.,10.4){.}
\put(81.,10.3){.}
\put(82.,10.2){.}
\put(83.,10.1){.}
\put(84.,10.1){.}
\put(85.,10.1){.}
\put(86.,10.){.}
\put(87.,10.){.}
\put(88.,10.){.}
\put(89.,10.){.}

\end{picture}

\vspace*{7mm}
\begin{picture}(120,30)

\put(110,13){(b)}

\put(60,16){\circle{2.}}
\put(62,16){\vector(1,0){6.}}
\put(70,15){$v_s$}

\put(0,14){$\rho(x)$}
\put(10,5){\vector(0,1){30}}
\put(59,0){$x$}
\put(15,5){\vector(1,0){90}}
\put(13,10){$\rho_{-}$}
\put(106,22){$\rho_{+}$}

\put(20.,10.){\circle*{.7}}
\put(21.,10.){\circle*{.7}}
\put(22.,10.){\circle*{.7}}
\put(23.,10.){\circle*{.7}}
\put(24.,10.){\circle*{.7}}
\put(25.,10.){\circle*{.7}}
\put(26.,10.){\circle*{.7}}
\put(27.,10.){\circle*{.7}}
\put(28.,10.){\circle*{.7}}
\put(29.,10.){\circle*{.7}}
\put(30.,10.){\circle*{.7}}
\put(31.,10.){\circle*{.7}}
\put(32.,10.){\circle*{.7}}
\put(33.,10.){\circle*{.7}}
\put(34.,10.){\circle*{.7}}
\put(35.,10.){\circle*{.7}}
\put(36.,10.){\circle*{.7}}
\put(37.,10.){\circle*{.7}}
\put(38.,10.){\circle*{.7}}
\put(39.,10.){\circle*{.7}}
\put(40.,10.){\circle*{.7}}
\put(41.,10.){\circle*{.7}}
\put(42.,10.){\circle*{.7}}
\put(43.,10.){\circle*{.7}}
\put(44.,10.){\circle*{.7}}
\put(45.,10.){\circle*{.7}}
\put(46.,10.){\circle*{.7}}
\put(47.,10.){\circle*{.7}}
\put(48.,10.){\circle*{.7}}
\put(49.,10.){\circle*{.7}}
\put(50.,10.){\circle*{.7}}
\put(51.,10.){\circle*{.7}}
\put(52.,10.){\circle*{.7}}
\put(53.,10.){\circle*{.7}}
\put(54.,10.){\circle*{.7}}
\put(55.,10.){\circle*{.7}}
\put(56.,10.){\circle*{.7}}
\put(57.,10.1){\circle*{.7}}
\put(58.,10.5){\circle*{.7}}
\put(59.,12.1){\circle*{.7}}
\put(60.,16.){\circle*{.7}}
\put(61.,19.9){\circle*{.7}}
\put(62.,21.5){\circle*{.7}}
\put(63.,21.9){\circle*{.7}}
\put(64.,22.){\circle*{.7}}
\put(65.,22.){\circle*{.7}}
\put(66.,22.){\circle*{.7}}
\put(67.,22.){\circle*{.7}}
\put(68.,22.){\circle*{.7}}
\put(69.,22.){\circle*{.7}}
\put(70.,22.){\circle*{.7}}
\put(71.,22.){\circle*{.7}}
\put(72.,22.){\circle*{.7}}
\put(73.,22.){\circle*{.7}}
\put(74.,22.){\circle*{.7}}
\put(75.,22.){\circle*{.7}}
\put(76.,22.){\circle*{.7}}
\put(77.,22.){\circle*{.7}}
\put(78.,22.){\circle*{.7}}
\put(79.,22.){\circle*{.7}}
\put(80.,22.){\circle*{.7}}
\put(81.,22.){\circle*{.7}}
\put(82.,22.){\circle*{.7}}
\put(83.,22.){\circle*{.7}}
\put(84.,22.){\circle*{.7}}
\put(85.,22.){\circle*{.7}}
\put(86.,22.){\circle*{.7}}
\put(87.,22.){\circle*{.7}}
\put(88.,22.){\circle*{.7}}
\put(89.,22.){\circle*{.7}}
\put(90.,22.){\circle*{.7}}
\put(91.,22.){\circle*{.7}}
\put(92.,22.){\circle*{.7}}
\put(93.,22.){\circle*{.7}}
\put(94.,22.){\circle*{.7}}
\put(95.,22.){\circle*{.7}}
\put(96.,22.){\circle*{.7}}
\put(97.,22.){\circle*{.7}}
\put(98.,22.){\circle*{.7}}
\put(99.,22.){\circle*{.7}}
\put(100.,22.){\circle*{.7}}
\put(101.,22.){\circle*{.7}}
\put(102.,22.){\circle*{.7}}
\put(103.,22.){\circle*{.7}}
\put(104.,22.){\circle*{.7}}

\end{picture}
\end{center}
\caption{\small (a)
Diffusive spreading of a density perturbation in the stationary state
at two times $t_2 > t_1$. The collective velocity describes the motion
of the center of mass of the perturbation. (b) Motion of a shock. To the left
of the domain wall particles are distributed
homogeneously with an average density $\rho_{-}$. To the
right of the domain wall the background density is $\rho_{+} > \rho_{-}$}
\label{v}
\end{figure}

The collective velocity is the velocity of the center of mass of a local
perturbation in a homogeneous stationary background (Fig.~\ref{v}a).
It is positive for background density $\rho < \rho^\ast$, but becomes
negative for $\rho > \rho^\ast$. In this case the perturbation creates
a back-moving traffic jam, which leads to a negative center-of-mass velocity,
even though all individual particles move with positive velocity.
In terms of traffic flow one might think of such a perturbation as being a
car which has just entered a major throughway from some side road.

The shock velocity describes the motion of the shock which performs,
due to fluctuations, a biased random walk with velocity $v_s$ (Fig.~\ref{v}b).
If the incoming current $j_-$ exceeds the outgoing current $j_{+}$, the
shock velocity is negative. This is analogous to the back-moving shock of a
traffic jam for sufficiently high incoming traffic flow.

To get an intuitive understanding how these velocities
determine the stationary phase diagram of a driven system coupled to
boundary reservoirs let us assume $\rho_{\pm}=\rho_{R,L}$.
We consider first $\rho_{+} > \rho^\ast$
where $\rho^\ast$ is the density where the current takes its maximal
value $j^\ast$. To make the argument more transparent we also assume that
particles hop only to the right and that initially the lattice is empty.
Because of ergodicity the stationary distribution is independent of the
initial state and hence this assumption involves no loss of generality.
Consider now the time evolution of the averaged density profile in  a
large system, starting from the empty lattice. We start the discussion
by assuming the left boundary density to be very low.

(i) As time proceeds, particles from the left reservoir will enter the system
and (possibly after some distance describing the non-universal
boundary layer) create a region of constant density $\rho_{-}$. This region
decays to the right to zero, because after a finite time the rightmost
particles will have traveled only a finite distance. Eventually however,
after a time which is of the order of system size, the rightmost particles
will hit the right boundary with the reservoir of density $\rho_{+}$. This
reservoirs makes it more difficult for particles to travel further and hence
creates a little traffic jam. The result is a shock profile, with shock
densities $\rho_{-}$ on the left and $\rho_{+}$ to the right resp. like
in Fig.~\ref{v}b. [For the sake of the argument one could have chosen such
an initial state. The reason for choosing an empty initial state becomes
clear below.] The decisive question is now how this shock profile evolves in
time. According to (\ref{3-2}) the shock velocity under the circumstances
described here is positive, simply because the incoming current of particles
$j_{-}$ is less than the outgoing current $j_{+}$ for sufficiently small left
boundary density. Hence, even though a shock forms by fluctuations, it
has an {\em average} drift towards the right boundary. Hence the system remains 
in the low density (LD) regime with bulk density $\rho = \rho_{-} < \rho^\ast$.

(ii) The situation changes when the left boundary density takes a value
such that the incoming current $j_{-}$ equals the outgoing current $j_{+}$.
In this case the shock velocity vanishes and the shock performs an unbiased
random walk over the lattice. Hence the density profile may be regarded
as being composed of two stationary domains with densities $\rho_{-}$ and
$\rho_{+}$, separated by a ``domain wall'' which is the sharp transitional
region of the shock. Since the shock motion is unbiased, the stationary
probability of finding the shock is constant in space. This leads to
an equal superposition of shock profiles and hence to a linearly increasing
stationary density profile. In analogy to first order equilibrium phase
transitions where a domain wall separates regions of coexisting
equilibrium regimes, we call the line defined by $j_{+}=j_{-}$ and
$\rho_{-} < \rho^\ast$, $\rho_{+} > \rho^\ast$, a first order phase transition
line.

(iii) This line marks the transition to a high-density phase (HD)
with bulk density $\rho = \rho_{+} > \rho^\ast$ since for even higher
left boundary density the incoming current
into the shock exceeds the outgoing current, and according to (\ref{3-2})
the shock moves to the left. This leaves the bulk in the high-density
regime determined by the coupling to the right boundary reservoir.
At the first-order transition line the stationary bulk density
is discontinuous, it jumps from $\rho_{-}$ to $\rho_{+}$.

Next we consider the case of low right boundary density $\rho_{+} < \rho^\ast$.
For definiteness we choose $\rho_{+} = 0$. The essential part of the
following discussion, the explanation of the occurrence of a continuous
phase transition, is unaffected by this choice.
Again we start the discussion with the empty lattice.

(i) As argued above, after some initial time the system will have filled up to
a bulk density $\rho = \rho_{-}$. Since $\rho_{+}=0$, particles hitting the
right
boundary can leave the system without creating a shock. As a result,
the system is in the low-density phase. [For $\rho_{+} > 0$ a shock could form,
but would have positive velocity under the circumstances considered here.
Hence also in this case the system is in the low-density phase.] Now we
examine the seemingly trivial reason {\em why}
an increase in the left boundary density leads to an increase in the
bulk density. Above we simply claimed this to be true on the basis
of plausibility. Here we support this
claim with an argument that becomes important below. Suppose we create a
little perturbation at the left boundary by injecting an extra particle
(on top of those particles that are injected anyway from the reservoir).
This creates a perturbation which, by definition of the collective velocity,
travels with $v_c > 0$ into the bulk and leads to a local increase of the
density, moving away from the boundary and spreading out as time goes on. 
The collective velocity is positive, since by assumption we have a background 
density $\rho_{-} < \rho^\ast$ and hence the current as a function of the 
density has positive slope. The point is that
maintaining such a perturbation (which corresponds to increasing the
left reservoir density permanently) leads to a permanent additional flow
of particles into the bulk and hence to the anticipated increase of 
the stationary bulk density.

(ii) It is now clear that this argument holds only as long as
$\rho_{-} < \rho^\ast$. Assume now $\rho_{-}=\rho^\ast$. Following the
reasoning
above this results in a maximal-current bulk density $\rho^\ast$.
However, if the left reservoir density increases beyond the
maximal current density, the collective velocity becomes negative. No extra
particles flow into the system, which therefore remains in its bulk at the
maximal current density  $\rho^\ast$. The system is now in the maximal current
phase for all $\rho_{-} > \rho^\ast$ and $\rho_{+} < \rho^\ast$. This
transition is continuous, the bulk density approaches $\rho^\ast$ smoothly
from below. Intuitively this phenomenon may be understood as ``overfeeding''
\cite{Schu93b}. The injected
particles act as blockages for further incoming particles, leading to a
back-moving traffic jam at the origin. This increased density at the origin
blocks further injection attempts and prevents the actual increase of the
current.

{}Finally, using similar arguments, one can show that the transition from the
high-density phase to the maximal current phase is also continuous.

To summarize, the theory predicts a first-order transition along the
line defined by $j_{+}=j_{-}$ and $\rho_{-} < \rho^\ast$, $\rho_{+} >
\rho^\ast$
where the stationary bulk density jumps from $\rho_{-}$ (low density phase LD)
to $\rho_{+}$ (high density phase HD). On the phase transition line the
stationary density is linearly increasing from $\rho_{-}$ to $\rho_{+}$.
{}From both phases there is a continuous phase transition to the maximal
current phase MC defined by $\rho_{-} > \rho^\ast$ and $\rho_{+} < \rho^\ast$.
In this phase the bulk density takes the maximal current value
$\rho =\rho^\ast$. These rules are encoded in an extremal principle
for the current 
\be
 j = \left\{ \ba{l} \displaystyle
\max_{\rho \in [\rho_R,\rho_L]} j(\rho) \quad \mbox{ for } \rho_L>\rho_R \\
\displaystyle
\min_{\rho \in [\rho_L,\rho_R]} j(\rho) \quad \mbox{ for } \rho_L<\rho_R.
\ea \right.
\ee
derived in \cite{Popk99a}. It is worthwhile pointing out that
this dynamical theory explains in mesoscopic terms the predictions one would 
obtain by viewing the system from a coarse-grained hydrodynamic view-point 
\cite{Krug91a}. At the same time, verification of this scenario suggests
the validity of the hydrodynamic approach to the description of the
large-scale dynamics of the particle system \cite{Spoh91} by using the exact 
current-density relation.

\subsection{Coupling to boundary reservoirs}

To verify this scenario which correctly describes the exactly solvable
ASEP with open boundaries one has to investigate our model in terms of the
effective boundary densities $\rho_{-},\rho_{+}$. Given two reservoir
densities
there is no general recipe how to eliminate the non-universal boundary
effects that result in the effective boundary densities $\rho_{-},\rho_{+}$.
Hence these quantities are not easy to control. Ideally, for purposes
of theoretical investigation, one would like to construct an injection and
absorption mechanism which leads to a
constant density profile for a semi-infinite system so that
$\rho_L = \rho_{-}$ and $\rho_R = \rho_{+}$, i.e. the effective
boundary densities are identical to the actual control parameters of the
model.

Here we choose an injection mechanism where the particles on the lattice
interact with the reservoir particles in the same way as among each other.
We define two injection rates from the reservoir at the left boundary:
\begin{itemize}
\item injection at site 1 if site 2 is occupied\\
$|\emptyset A \; \to \; |AA$ with rate $\alpha_1$
\item injection at site 1 if site 2 is empty\\
$|\emptyset \emptyset \; \to \; |A \emptyset$  with rate $\alpha_2$
\end{itemize}
and two new hopping rates at the right boundary:
\begin{itemize}
\item hopping from site $N-1$ to site $N$\\
$A \emptyset | \; \to \; \emptyset A|$ with rate $\beta_1$
\item hopping out from site $N$ (absorption)\\
$A | \; \to \; \emptyset |$ with rate $\beta_2$.
\end{itemize}
These four hopping rates are those that would be affected by the
interaction with particles in the reservoirs. The rates $\alpha_i$
($\beta_i$ resp.) have now to be determined as functions of the
left (right) reservoir density. This can be illustrated e.g.
for the injection process with rate $\alpha_1$.
We imagine the reservoir to include a site 0 of the chain.
The injection rate into the first site is defined by the (stationary)
average occupation $\rho_L$ of the imaginary site 0,
but with the condition that the first site is empty and the second
site is occupied. Considering the zeroth, the first and the second site as
three neighboring sites of an infinite chain,
this conditional probability can be expressed readily as
correlations in the stationary state of an infinite chain.
Thus we find $\alpha_1 = q \exval{101}/\exval{01}$ where expectation values
like $\exval{101} = \exval{n_i (1-n_{i+1})n_{i+2}}$ are calculated in the
thermodynamic limit of the distribution (\ref{2-1}) with density
$\rho = \rho_{L}$. The case of $\alpha_2$ is entirely similar and one finds
$\alpha_2 = r \exval{100}/\exval{00}$.

{}For the calculation of the right
boundary rates we note that the probability of jumping from the site
$N-1$ to the site $N$ is
affected by the average occupation of the imaginary reservoir site $N+1$.
With the jump condition that site $N-1$ is occupied and site $N$ is empty
one finds $\beta_1 = (r \exval{100} + q \exval{101})/\exval{10}$.
The case of $\beta_2$ is similar but one has to take into account
the conditional probability of the occupation of two
imaginary reservoir sites $N+1$ and $N+2$. This gives
$\beta_2 = (r \exval{100} + q \exval{101})/\exval{1}$.
One has to determine these correlations in an infinite chain with
density $\rho_{R}$. These rates can be expressed
as a function of the density through the form of the current and we find:
\bea
\label{3-3a}
\alpha_1 &=& q {\exval{101} \over \exval{01}} = 
	q  { \exval{10}_{\rho_{L}} \over 1-\rho_L}  \\
\label{3-3b}
\alpha_2 &=& r {\exval{100} \over \exval{00}} =
	r  { \exval{10}_{\rho_{L}} \over 1-\rho_L}  \\
\label{3-3c}
\beta_1 &=& {r \exval{100} + q \exval{101} \over \exval{10}} =
	{ j(\rho_{R}) \over \exval{10}_{\rho_{R}} } \\
\label{3-3d}
\beta_2 &=& {r \exval{100} + q \exval{101} \over \exval{1}} =
	{j(\rho_{R}) \over \rho_{R}}
\eea
with
\be
\exval{10}_\rho = \exval{01}_\rho = 
   (1-\rho) \left( 1 - {j(\rho) \over r\rho}\right)
\ee
{}For $\rho_{L}=1$ we use the limiting values  $\alpha_1=q$,
$\alpha_2=r$, and for $\rho_{R}=0$ we use $\beta_1=\beta_2=r$.
Together with the bulk hopping rates $r,q$ the dynamics of the model
is now completely defined.

Before discussing the full phase diagram we consider the case of equal
boundary densities $\rho = \rho_{R} = \rho_{L}$. Somewhat surprisingly, it
turns out that we can actually obtain the full stationary distribution of the
process:
\bel{3-4}
P^\ast(\{n\})=
 {1-\rho \over \lambda_1^{N-1}}
 \left({q \over r}\right)^{\sum_{i=1}^{N-1}n_i n_{i+1}}
 z^{\sum_{i=2}^{N-1}n_i} (\lambda_1-1)^{n_1+n_N}
\ee
with the eigenvalue $\lambda_1$ of the transfer matrix of the one-dimensional
Ising model and the ``fugacity'' $z=e^{-\beta h}$ (Appendix A).
Stationarity of this distribution can be proved by writing the time
evolution operator of the process in a quantum Hamiltonian formalism
(see appendix B). Like in
the periodic case, the exact stationary non-equilibrium distribution of the
open system with equal boundary densities is the equilibrium distribution of
an Ising chain of length $N$, but with boundary fields $g(n_1+n_N)$ rather than
with the Ising coupling $n_N n_1$.

Moreover, with the transfer matrix formulation
of the Ising model it is straightforward to show that the density profile
is constant. The theoretical scenario described above
then suggests the identification $\rho_{-} = \rho_L$ for $\rho_L<\rho^\ast$ and
$\rho_{+} = \rho_R$ for $\rho_R>\rho^\ast$. For $\rho_L>\rho^\ast$ or
$\rho_R<\rho^\ast$ respectively constant profiles are not stable with
respect to fluctuations and the relationship between the reservoir
densities and the effective boundary densities may be more subtle (see below).

\section{Phase diagram for repulsive interaction}
\label{rep}

Having defined the model and given the theoretical background we are now
set to investigate the phase diagram of the system. We consider first
the traffic flow scenario (repulsive interaction).

\subsection{Mean field}

Unfortunately the exact stationary distribution of the open system
for different boundary densities does not have a simple form. However,
also the density in the open system satisfies a continuity equation
of the form (\ref{2-1a}), with the boundary currents
\bea
& & j_0 = \alpha_1 \exval{(1-n_1)n_2} + \alpha_2 \exval{(1-n_1)(1-n_2)}
 \label{boundcur1} \\
& & j_{N-1} = \beta_1 \exval{n_{N-1}(1-n_N)} \\
& & j_N = \beta_2 \exval{n_N}
\label{boundcurN}
\eea
Using also the expression (\ref{2-2}) for the bulk current
one can obtain the stationary density profile in a mean field approximation
by neglecting the correlations in the expectation values.
Stationarity implies equal current everywhere in the lattice,
$\exval{j_i} = j$. This gives rise to the bulk recursion relation
\bel{recuri}
\rho_i = {j \over (1-\rho_{i+1})\left( q\rho_{i+2} +
 	r(1-\rho_{i+2})\right)}
\ee
for the density profile. The boundary values of the recursion are determined
by the boundary currents (\ref{boundcur1}) - (\ref{boundcurN}).

Because of the interaction with the boundary
there are two possibilities to perform a mean-field analysis. The most
straightforward choice would be to choose the boundaries rates
(\ref{3-3a}) -  (\ref{3-3d}) and analyze the mean field phase diagram in terms
of the reservoir densities $\rho_{R,L}$. For a given choice of current one can
draw a density profile and read off the various phases. However, it turns out
that within such an approximation scheme a constant density profile cannot be 
achieved even if $\rho_R = \rho_L$
and that the boundary densities are not equal to the reservoir densities.
As a result, the mean field phase diagram obtained in this way
differs considerably from the theoretical expectation. Within this
approximation
the MC phase disappears for $r/q$ smaller than $\approx 0.6$.

{}From a theoretical point of view it is more natural to neglect the
correlations even in the expressions of the boundary rates in terms of
the Ising expectation values. This gives rise to
simple expressions of the boundary rates in terms of the reservoir
densities. Using the boundary conditions $\rho_0 = \rho_{L}$ and
$\rho_{N+1} = \rho_{N+2} = \rho_{R}$ extends the validity of the recursion
(\ref{recuri}) to the range $0 \leq i \leq N$.
In this way $\rho_N$ can be expressed as a function of $\rho_R$ and $j$
and this was the reason of choosing the direction of the recursion 
from the right to the left.
It is easy to see that for a semi-infinite system there are the constant
solutions $\rho_i = \rho_L$ and $\rho_i = \rho_R$ resp.
with the mean field current
\be
j^{MF}(\rho) = \rho (1-\rho) \left( q\rho + r(1-\rho)\right)
\ee
Like in the exact
solution this is also a solution of the finite system with equal boundary
densities. Therefore we shall discuss this mean-field approach in some
more detail.

{}For general boundary densities we did not find a solution of the 
recursion (\ref{recuri}) in closed form, but it is easily solved numerically
on a computer. It is sufficient to choose a density $\rho_{R}$ and a current 
$j$ and apply the recursion relation for drawing the density profile.
If any of $\rho_i$'s is smaller than 0 or greater than 1 then there
is no solution corresponding to these values of $\rho_{R}$ and $j$
at this system size since any physical solution has to satisfy
$0 \leq \rho_i \leq 1$ for all $1\leq i\leq N$. For fixed $\rho_{R},\rho_{L}$
and length $N$ this requirement fixes the current $j$.
In this way one can map out the whole phase diagram,
finding the corresponding current for all values of
$\rho_{L}$ and $\rho_{R}$. For sufficiently large $N$ the size dependence
is negligibly small.

\begin{figure}[htb]
\centerline{
\epsfxsize=0.4\textwidth
\epsfbox{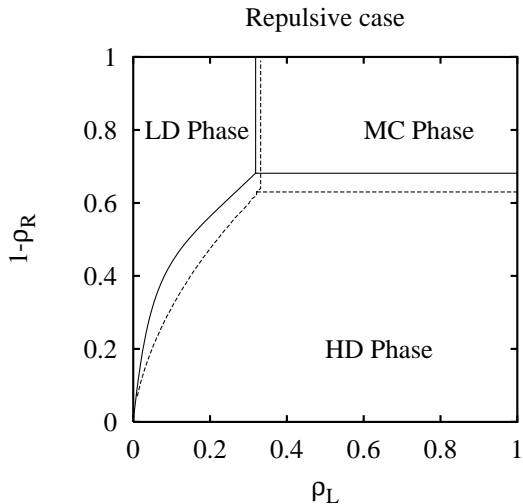}}
\caption{Mean field phase diagram (broken curves) and theoretical phase
transition lines (full curves) for repulsive interaction.}
\label{phase1mf}
\end{figure}

We have chosen $r=1.0$ and $q=0.1$.
The low- (LD) and the high-density (HD) phases resp. are easy to distinguish as
the bulk density is equal to one of the end densities. This allows us to
identify $\rho_{-} = \rho_L$ and $\rho_{+} = \rho_R$.
At the other end oscillatory behavior is observed.
This does not happen in the usual ASEP with open boundaries
\cite{Schu93b,Derr93a}, but has been observed in an ASEP with particles
covering more than one lattice site \cite{MacD68}. The maximal mean field 
current accessible at given
parameters $r$ and $q$ and the corresponding bulk density define
the maximal current phase MC. The location of these phases in the parameter
space differs from those obtained from theoretical scenario reviewed above
(Fig.~\ref{phase1mf}) by using the exact value (\ref{2-3}) of the current.
{}Further analysis
shows that the discrepancy increases with increasing repulsion.

\subsection{Monte Carlo results}
\label{MC}

In order to check the phase diagram based on the theory of boundary
induced phase transition described above
we have performed Monte Carlo simulations of the model.

The boundary rates are determined as explained above to simulate given left
and right densities. During one MC step one of $N+1$ sites is chosen randomly
($N$ is the size of the system and the +1 is for the jumping into the system)
and if there is a particle then it can hop with a probability given by
the hopping rates.
The initial configuration was an empty lattice.
The required time to reach the stationary state for given system size
can be determined by investigating the time dependence of the current
and the bulk density (the density in the middle of the system).
The order parameter which is the stationary value of the bulk density
obtained as an average over $10^7$ MC steps is the best indicator of the phase
transition lines. At the first order phase transition
between the high and the low density phases
it has a pronounced jump from the left to the right density.
A system size of 1000 is sufficiently large to localize
the phase transition line. The transition into the maximal current phase is
continuous, only the derivative of the order parameter has a (jump)
discontinuity. In order to localize this phase transition line
one needs larger systems of size 5000.

\begin{figure}[htb]
\centerline{
\epsfxsize=0.4\textwidth
\epsfbox{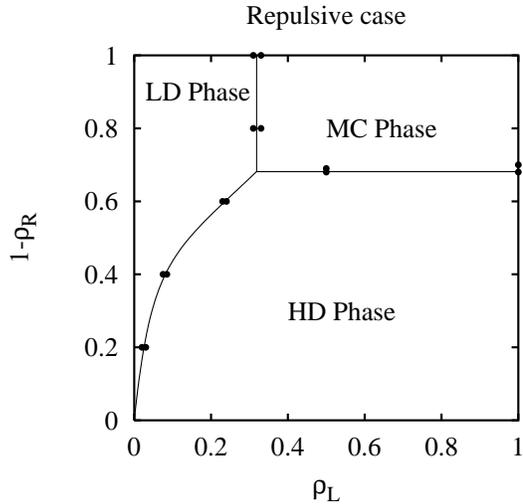}}
\caption{Monte Carlo data and theoretical phase
transition lines (full curves) for repulsive interaction.
The numerical error bars are indicated by full circles.}
\label{phase1}
\end{figure}

In a given phase the simulated bulk density depends on the boundary
densities the same way as it is predicted by the theory, namely
it equals to the left (right) density in the LD (HD) phase,
and it is a well defined constant value in the MC phase.
At the phase boundary the crossover between two kinds of behavior
of the bulk density can be localized within an error
shown the figures (). The Monte Carlo
results are seen to be in agreement with the phase diagram derived from the
theory of boundary-induced phase transitions (Fig.~\ref{phase1}).
The mean-field phase transition lines are significantly outside the numerical
error bars except on part of the transition line between LD and MC phase.

\section{Attractive interaction}
\label{att}

\begin{figure}[htb]
\centerline{
\epsfxsize=0.4\textwidth
\epsfbox{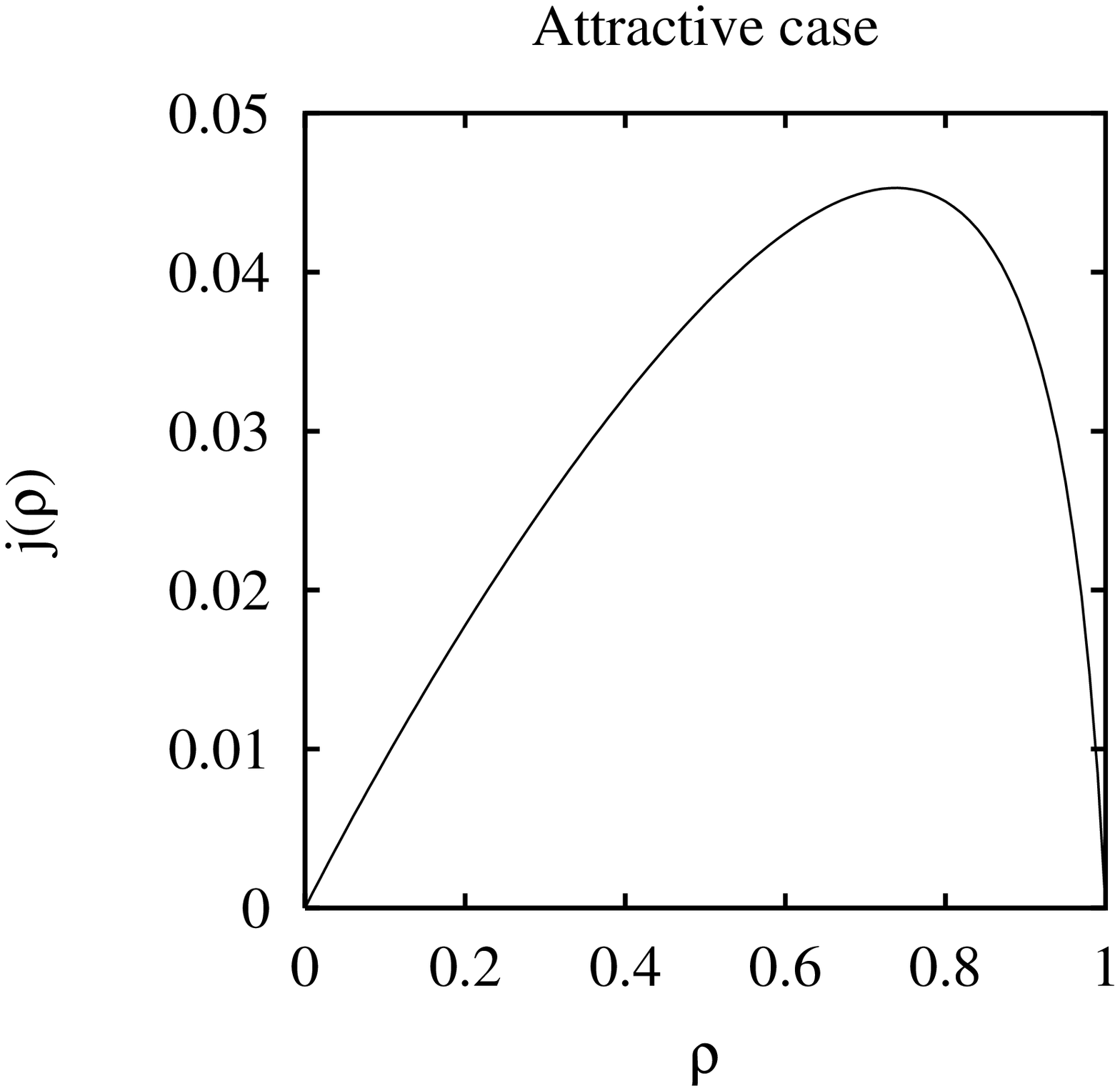}}
\caption{Exact stationary current $j$ as a function of the density $\rho$
for $q=1,\, r=0.1$ (repulsive interaction).}
\label{curratt}
\end{figure}

Attractive (ferromagnetic) interaction leads to bulk particle domains as in the
one-dimensional Ising model. Dynamically this is qualitatively understandable 
since because of $q>r$ particles tend to form clusters. One expects
a shift of the maximal-current density $\rho^\ast$ to higher densities.
This can indeed be shown by calculating the derivative of the exact
current (\ref{2-3}). The full current-density relation for $q=1$, $r=0.1$
is shown in Fig.~\ref{curratt}.

Analysis of the model with open boundaries proceeds in the same way as in the 
repulsive case. With $r=0.1$ and $q=1.0$ the mean
field approximation yields the three phases discussed above, with the
phase transition lines differing substantially from those expected
theoretically if one identifies $\rho_{-} = \rho_L$ and
$\rho_{+} = \rho_R$ as suggested by the density profiles
in the low and high density phases respectively.

\begin{figure}[htb]
\centerline{
\epsfxsize=0.4\textwidth
\epsfbox{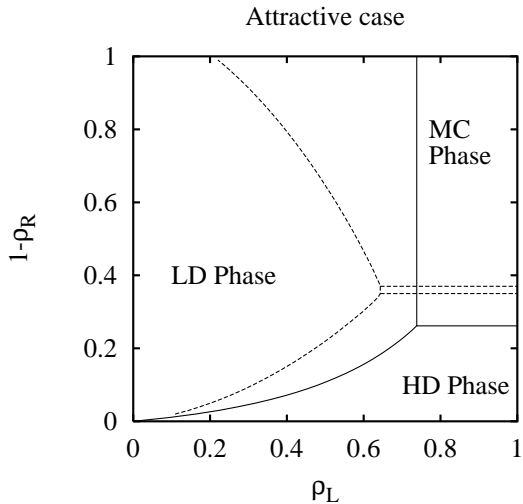}}
\caption{Mean field phase diagram (broken curves) and theoretical phase
transition lines (full curves) for repulsive interaction. The narrow area above
the HD phase is the mean field MC phase. The fourth phase obtained from
mean field covers the large area in the upper right part of the phase diagram.}
\label{phase2mf}
\end{figure}

As additional feature the mean field theory predicts a further phase transition
located at the large $\rho_{L}$ and small $\rho_{R}$ corner of the phase
diagram  (Fig.~\ref{phase2mf}). In this ``fourth phase'' the bulk density is
larger than the density corresponding to the maximal current.
The area of this phase increases with increasing attraction,
and the area of MC phase decreases but does not disappear.

\begin{figure}[htb]
\centerline{
\epsfxsize=0.4\textwidth
\epsfbox{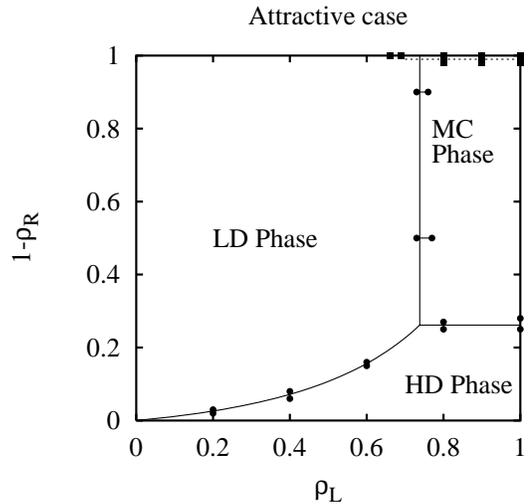}}
\caption{Monte Carlo data and theoretical phase
transition lines (full curves) for repulsive interaction.
The numerical error bars are indicated by full circles. The reentrance
phase transition to the fourth (HD) phase (upper right corner of the phase
diagram) is marked by the broken curve interpolating between the Monte Carlo 
points.
}
\label{phase2}
\end{figure}

The phase diagram obtained from Monte-Carlo simulation has the phase 
transitions expected from the theoretical prediction, but also shows that the 
fourth phase exists in a small neighborhood of the ($\rho_{+}=0$, 
$\rho_{-}>\rho^*$) line and hence is not an artefact of the mean-field
approximation. The location of the phase transition lines is rather
different compared to mean field (Fig.~\ref{phase2}). The transition to this 
phase from the LD phase is first order and from the maximal current phase it 
is second order. In this sense the fourth phase is like the high-density
phase, however, the bulk density differs from both the boundary densities
$\rho_{R,L}$. Within the theory of boundary-induced phase
transitions this high density phase can be understood by recalling the
non-universal relationship between real boundary density and effective
boundary density (in terms of which the theory is formulated). While for
repulsive interaction and the choice of coupling mechanism used here these
two quantities may be identified, the connection is apparently more
complicated for attractive interaction if the (real) boundary densities
become sufficiently high (left boundary) and low (right boundary) respectively.
An adequate explanation of this boundary phenomenon which leads to
a reentrance transition into the high-density phase is not available.
In order to rule out the possibility of a finite-size effect we performed
simulations with different system sizes. Using N=500, 1000 and 5000 resp.
we found no indication that the reentrance transition would disappear for 
large enough systems.

\section{Conclusions}
\label{conc}

Our analysis of the totally asymmetric exclusion model with next-nearest-neighbor
interaction consists of two parts. First we considered the periodic system
in order study the bulk properties of the steady state. With a view on
traffic flow modelling we remark that the observations (i) - (iv) listed above
are consistent with the results obtained here. The current-density relation 
becomes asymmetric (Fig.~\ref{AS98F-1}) in a way which is closer
to real traffic data as the symmetric relation $j = \rho (1-\rho)$ for the
ASEP with $r=q=1$. There is no discontinuity in the derivative at the
maximal-current density $\rho^\ast$, in agreement with the
arguments given above. 
As more far-reaching conclusion we note that the exact result (\ref{2-1}) sheds 
light on the so far somewhat unclear relationship between
the updating mechanism and the occurrence of correlations.
In order to obtain correlations 
in traffic flow models it is not essential to use a discrete-time updating 
mechanisms. Since other discrete-time models have uncorrelated stationary
states \cite{Raje97} it appears that correlations have their physical origin
in speed-reduction rather than in the nature of the updating scheme.

In the second part we investigated the steady-state selection in the open
system. It turns out that the theoretical scenario based on the interplay
of shocks and overfeeding correctly describes the phase diagram in terms of
effective boundary densities. For repulsive interactions this strengthens the 
case for a maximal current phase in real traffic. For attractive 
interaction there is, however, a 
surprising reentrance phase transition to the high density phase which 
originates in the so far poorly understood relationship between actual and 
effective boundary densities. Apparently these two are not always monotic 
functions of each other as one would naively expect. For a deeper understanding
the next step to be done is the analysis of the density profiles close
the reentrance phase transition lines in order analyze whether universal
properties of the usual transition lines \cite{Krug91a,Oerd98,Kolo98}
can be observed or not.

T.A. thanks for partial support by 
the Hungarian Academy of Sciences (Grant OTKA T 029792).
G.M.S. thanks K. Klauck, V. Popkov, L. Santen and A. Schadschneider
for illuminating discussions on traffic flow.

\appendix
\setcounter{section}{0}
\renewcommand{\theequation}{\Alph{section}.\arabic{equation}}

\section{Transfer matrix for the one-dimensional Ising model}
\setcounter{equation}{0}

The energy of the one-dimensional Ising model with the classical
spin variables $s_i = \pm 1$ can be written in term of the variables
$n_i = (1-s_i)/2$ in the form $E = J \sum_i n_i n_{i+1}$. In this
interpretation
the Ising model is a classical lattice gas of $M = \sum_i n_i$
hard-core particles with nearest neighbor interaction. Using standard
techniques \cite{Baxt82} one can express the grand canonical partition
function $Z = \sum_{config} z^M e^{- \beta E}$ at inverse temperature
$\beta = 1/kT$ in terms of the eigenvalues of the transfer matrix
\be
T = \left( \ba{cc} 1 & \sqrt{z} \\
                   \sqrt{z} & z e^{-\beta}
                                        \ea \right)
\ee
where, for our model, $e^{-\beta J} = q/r$ and we define $z=e^{-\beta h}$.
In the spin interpretation of the model $h$ plays the role of a magnetic field.

The eigenvalues of the transfer matrix are
\be
 \lambda_{1,2} = {1\over2}(1+{q\over r}z) \pm \sqrt{{1\over4}
 (1+{q\over r}z)^2 + z(1-{q\over r})}
\ee
where we choose $\lambda_{1}$ to the positive sign.
For a system with $N$ sites and periodic boundaries one has $Z = \mbox{Tr} T^N
= \lambda_1^N + \lambda_2^N$. The equilibrium distribution of the Ising
model is the stationary distribution of the particle hopping model. This
does {\em not} mean that the particle hopping model reaches thermal equilibrium
at long times, since the stationary distribution does not satisfy detailed
balance with respect to the dynamics of the model. The non-equilibrium
nature of the steady state results in a non-vanishing stationary particle
current. We stress that completely different dynamical models may have the
same stationary distribution, see e.g. the list of models given in
\cite{Katz84}. Another 
non-equilibrium example with the same Ising distribution is the
discrete-time ASEP with parallel update \cite{Yagu86}. Even though
the distribution is the same and hence all particle correlations are the
same, the stationary current of this model \cite{Scha93,Schr95}
(which is defined via the dynamics by the continuity equation) is
different from the current (\ref{2-2}) in our model.

In order to calculate expectation values in the transfer matrix formalism
we define the diagonal matrix
\be
n = \left( \ba{cc} 0 & 0 \\
                   0 & 1
                           \ea \right).
\ee
The density is then given by $\rho =
\mbox{Tr} (n T^N) / Z_N$ and a two-point correlation function is given
by $\exval{n_{k} n_l} = \mbox{Tr} (n T^{k-l} n T^{N-k+l})$. Higher order
correlators are calculated analogously. By diagonalizing $T$ one obtains
the expression
\be
 \rho = {1-\lambda_1 \over \lambda_2-\lambda_1}.
\ee
for the particle density in terms of the eigenvalues of $T$.
This relation may be used to express the current (\ref{2-2})
as a function of the particle
density. In the thermodynamic limit $N \to \infty$ one obtains (\ref{2-3}).

We note that expectation values for the distribution (\ref{3-4}) are
calculated with the same transfer matrix as in the periodic case. However,
instead of taking a trace, one calculates a scalar product with suitably
chosen vectors which are determined by the boundary fields.

\section{Proof of stationarity for equal boundary densities}
\setcounter{equation}{0}

The proof of stationarity of the distribution (\ref{2-1}) for the periodic
system is given in Ref. \cite{Katz84}. An alternative, constructive proof can be
obtained by using translational invariance and taking the distance between
neighboring particles as stochastic variables. In this way the particle hopping
process turns into a zero-range process \cite{Spit70} for which the stationary
distribution is known. Reexpressing the stationary zero-range distribution
in terms of particle occupation numbers yields (\ref{2-1}).

Here we prove stationarity of the Ising distribution (\ref{3-4}) for the
open system coupled to reservoirs of equal density. We use a convenient
standard approach for the stochastic description of classical interacting
particle systems, known as ``quantum Hamiltonian formalism'' \cite{Schu00}.
The basic idea is to formulate the generator of the Markov process in terms
of a many-body quantum operator. For processes without exclusion one obtains
in this way a Fock space representation of the generator in terms of bosonic
creation and annilation operators \cite{Kada68,Doi76,Gras80}. For exclusion
processes with at most one particle per site the same strategy yields an
operator expressed in terms of Pauli-spin matrices
\cite{Schu00,Sigg77,Sand93,Alca94}.

We define the exclusion process with state space $X=\{0,1\}^N$ and
transition rates $w_{\underline{n}\to\underline{n}'}$ from state
$\underline{n}$ to $\underline{n}'$ in terms of a master equation
\be
\frac{d}{dt} P(\underline{n};t) = \sum_{\underline{n}'\in X}
\left[w_{\underline{n}'\to\underline{n}}P(\underline{n}';t) -
w_{\underline{n}\to\underline{n}'}P(\underline{n};t)\right]
\ee
for the
probability $P(\underline{n};t)$ of finding, at time $t$, a configuration
$\underline{n}$ of particles on a lattice of $N$ sites. Here $\underline{n}
= \{n_1, n_2, \dots , n_N\}$ where $n_i=0,1$ are the integer-valued particle
occupation numbers at site $i$. The master equation is a linear first order
DGL in the time-variable and therefore it is natural to write it in a vector
notation with the probabilities $P(\underline{n};t)$ as vector components.
We represent each of the possible particle configurations $\underline{n}$ by a
column vector $\ket{\underline{n}}$ which form a basis of a vector space
${\bf X}=({\bbbc}^2)^{\otimes N}$. The transposed vectors $\bra{\underline{n}}$
form a basis of the dual space and we define the usual scalar product
$\bra{\underline{n}}\,\underline{n}'\,\rangle =
\delta_{\underline{n},\underline{n}'}$. The probability distribution is now
represented by a state vector $| \, P(t)\, \rangle = \sum_{\underline{n} \in X}
P(\underline{n};t) \ket{\underline{n}}$ and one can write the master equation
in the form
\bel{A-1}
\frac{d}{dt} P(\underline{n};t) = - \langle \, \underline{n} \, |
H | \, P(t) \, \rangle
\ee
where the off-diagonal matrix elements of $H$ are the (negative)
transition rates between states and the diagonal entries are the inverse
of the exponentially distributed life times of the states.
In formal analogy to the quantum mechanical Schr\"odinger equation
we shall refer to $H$ as quantum Hamiltonian.
A state at time $t'=t_0 + t$ is given in terms of an initial
state at time $t_0$ by
\bel{A-2}
| \, P(t_0+t) \, \rangle = \mbox{e}^{-H t }
| \, P(t_0) \, \rangle.
\ee

We stress that the physicists notion ``quantum Hamiltonian'' for the matrix
$H$ is somewhat misleading in so far as $H$ is, in fact, the
generator of the Markov semigroup of the process, rather than
the Hamiltonian of an actual quantum system. This by now well-established
notion has its origin in the fact that for various stochastic processes
the generator $H$ is identical to the quantum Hamiltonian of
some well-known spin system.
In this context we would also like to point out that quantum mechanical
expectation values $\exval{A} \equiv \bra{\Psi} A \ket{\Psi}$ for an
observable $A$ are calculated in
a different way than probabilistic expectation values for a function
$F(\underline{n})$ of the stochastic variables $\eta$. In the quantum
Hamiltonian
formalism one writes $\exval{F} \equiv \sum_{\underline{n} \in X}
{}F(\underline{n})
P(\underline{n};t)
= \bra{s} F \ket{P(t)}$ with the  matrix $F = \sum_{\underline{n} \in X}
{}F(\underline{n})
\ket{\underline{n}} \bra{\underline{n}}$ and the summation vector
$\bra{s} = \sum_{\underline{n} \in X} \bra{\underline{n}}$ which performs the
average over all
possible final states of the stochastic time evolution.

{}For our considerations the expectation value
$\rho_k(t)=\bra{s} n_k \ket{P(t)}$ for the density at site $k$ is of
special interest. It is given by the projection operator $n_k$ which has
value 1 if there is a particle at site $k$ and 0 otherwise.
$m$-point density correlations are then given by the expression
$\bra{s} n_{k_1} \dots n_{k_m} \ket{P(t)}$. In this paper we study only
stationary expectation values. For the formal description of a stationary
probability distribution we use the transposed summation vector
$\ket{s} = \sum_{\underline{n} \in X} \ket{\underline{n}}$.
A general stationary measure $P^\ast(\underline{n})$ may then be written
in vector notation in the form $\ket{P^\ast} = e^{-\beta E(\underline{n})}
\ket{s}/Z_N$ with the configuration-dependent ``energy'' matrix
$E(\underline{n})$ and the ``partition function'' $Z_N = \bra{s}
e^{-\beta E(\underline{n})} \ket{s}$ which normalizes the measure
$\ket{\tilde{P}^\ast} = e^{-\beta E(\underline{n})}\ket{s}$. Notice that
in vector notation the expression $E(\underline{n})$ represents a diagonal
matrix with the energies as diagonal elements. Below we shall also
use the invertible diagonal matrix $P^\ast = e^{-\beta E(\underline{n})}$
which has the (unnormalized) stationary probabilities $P^\ast(\underline{n})$
as diagonal elements.

To obtain the quantum Hamiltonian for the time evolution of the interacting
ASEP with open boundaries
we note that one can represent any two-state particle system as a spin
system by identifying a particle (vacancy) on site $k$ with a spin-down
(up) state on this site. This allows for a representation of $H$
in terms of Pauli matrices where $n_k =( 1 - \sigma^z_k )/2$
projects on states with a particle on site $k$ and $v_k = 1-n_k$ is the
projector on vacancies. The off-diagonal matrices
$s^{\pm}_k = (\sigma^x_k \pm i \sigma^y_k)/2$
create ($s^-_k$) and annihilate ($s^+_k$) particles. We stress that
in the present context the ``spins'' are just convenient labels for
particle occupancies. Using this pseudospin formalism one finds
\bea
H & = &  \sum_{k=1}^{N-2}
(n_k v_{k+1} - s^+_ks^-_{k+1})(r n_{k+2} + q v_{k+2}) \nonumber \\
 \label{A-3}
 &  &  + (1-n_1-s^-_1)(\alpha_1 n_2 + \alpha_2 v_2) \\
 &  &  + \beta_1 (n_{N-1} v_N - s^+_{N-1}s^-_N) + \beta_2 (n_N - s^+_N).
\nonumber
\eea

Within this formalism proof of stationarity is now a straightforward
calculation, using simple expressions for the jumping rates at the ends:
\bea
\alpha_1 & = & \nonumber {qz \over \lambda_1(\lambda_1-1)} \\
\alpha_2 & = & \nonumber {rz \over \lambda_1(\lambda_1-1)} \\
\beta_1 & = & \nonumber {qz \over \lambda_1-1} \\
\beta_2 & = &  {qz \over \lambda_1} .
\eea
and assuming that the boundary densities are equal at the two end.
According to (\ref{A-2}) stationarity is equivalent to the eigenvector relation
$H \ket{P^\ast} = 0$ which in turn is equivalent to
\be
(P^\ast)^{-1} H \ket{\tilde{P}^\ast} =
(P^\ast)^{-1} H P^\ast \ket{s} = 0.
\ee
The diagonal similarity transformation of $H$ with $P^\ast$ leads to a sum of
transition matrices which act non-trivially on at most four neighboring sites.
To calculate the action of the non-diagonal parts on $\ket{s}$ we use
$s^+_k \ket{s}
= v_k \ket{s}$ and $s^-_k \ket{s} = n_k \ket{s}$. This leaves only diagonal
terms the sum of which vanishes identically. This proves stationarity of
the measure (\ref{3-4}).

\end{document}